\documentclass[aps,prl,twocolumn,showpacs,floatfix, nobibnotes, superscriptaddress, citesort]{revtex4-1}

\usepackage{mathrsfs}
\usepackage{graphicx,graphics,color,epsfig}
\usepackage{bm}
\usepackage{amsmath}
\usepackage{amssymb}
\usepackage{subfigure}
\usepackage{amsfonts}

\usepackage{amstext}
\usepackage{float}

\usepackage[dvipdfm,  
            pdfstartview=FitH,
            CJKbookmarks=true,
            bookmarksnumbered=true,
            bookmarksopen=true,
            colorlinks,
            pdfborder=001,
            linkcolor=black,
            anchorcolor=blue,
            citecolor=blue
            ]{hyperref}
\begin{document}

\title{Quantum phase transition of cold atoms in the bilayer hexagonal optical lattices}

\author{Wei Xie}

\affiliation{National Laboratory for Condensed Matter Physics,
Institute of Physics, Chinese Academy of Sciences, Beijing 100190,
China}
\affiliation{School of Electronics and Information Engineering, Sichuan University, Chengdu 610065, China}

\author{Qi-Hui Chen}

\affiliation{National Laboratory for Condensed Matter Physics,
Institute of Physics, Chinese Academy of Sciences, Beijing 100190,
China}
\affiliation{School of Electronics and Information Engineering, Sichuan University, Chengdu 610065, China}

\author{Wu-Ming Liu}

\affiliation{National Laboratory for Condensed Matter Physics,
Institute of Physics, Chinese Academy of Sciences, Beijing 100190,
China}

\begin{abstract}

We propose a scheme to investigate the quantum phase transition of cold atoms in the bilayer hexagonal optical lattices. Using the quantum Monte Carlo method, we calculate the ground state phase diagrams which contain an antiferromagnet, a solid, a superfluid, a fully polarized state and a supersolid. We find there is a supersolid emerging in a proper parameter space, where the diagonal long range order coexists with off-diagonal long range order. We show that the bilayer optical lattices can be realized by coupling two monolayer optical lattices and give an experimental protocol to observe those novel phenomena in the real experiments.

\end{abstract}


\maketitle

\begin{figure}[H]
\centering
\vspace{0pt}
\begin{minipage}[t]{0.5\textwidth}
\centering
\includegraphics[width=2.1in]{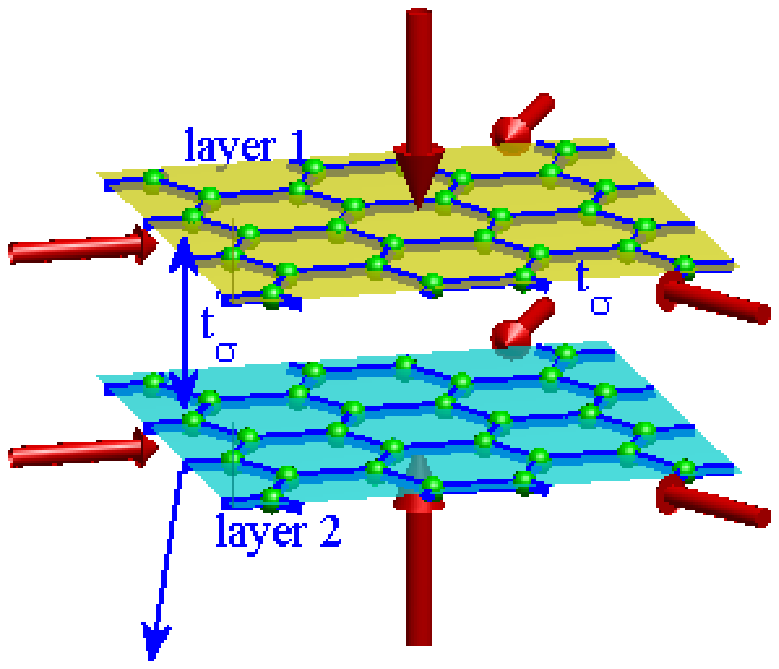}\\
\includegraphics[width=1.5in]{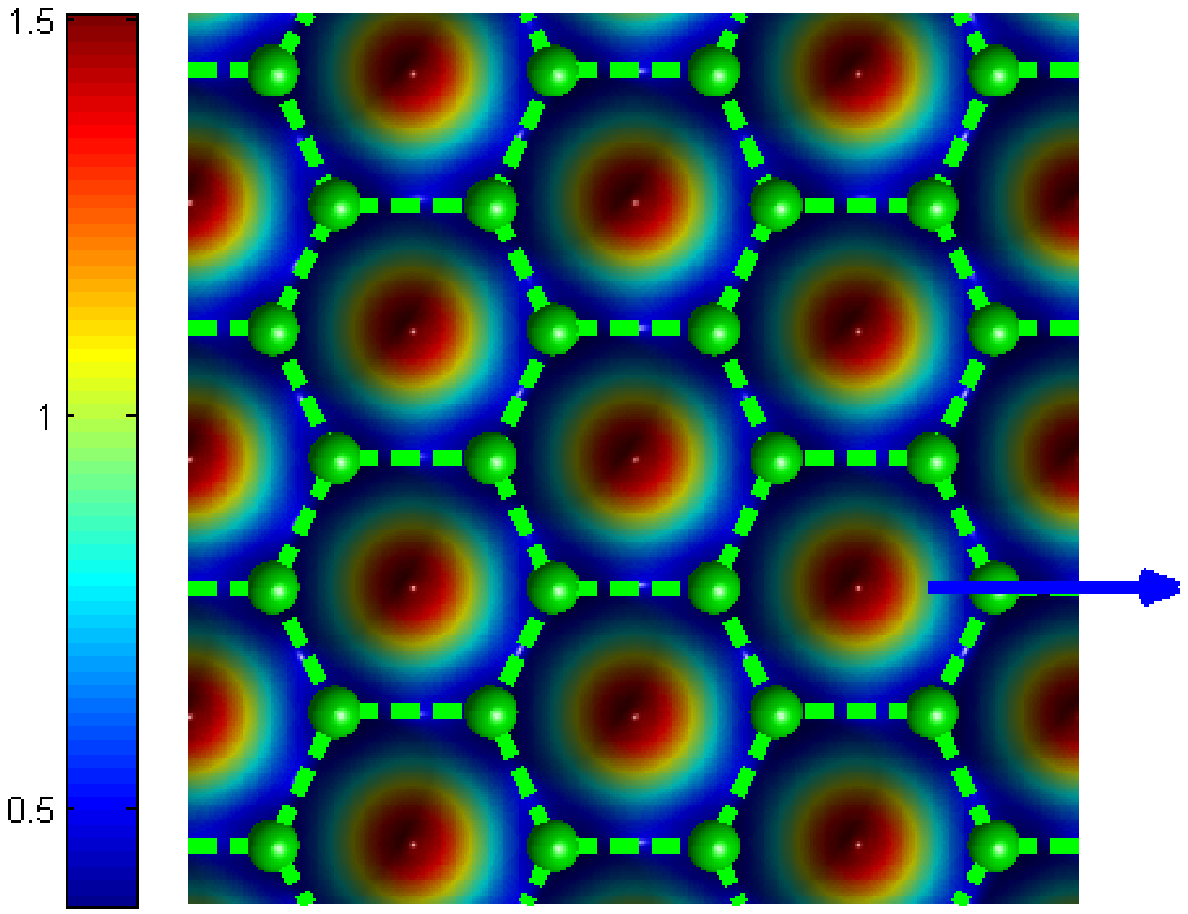}
\includegraphics[width=1.2in]{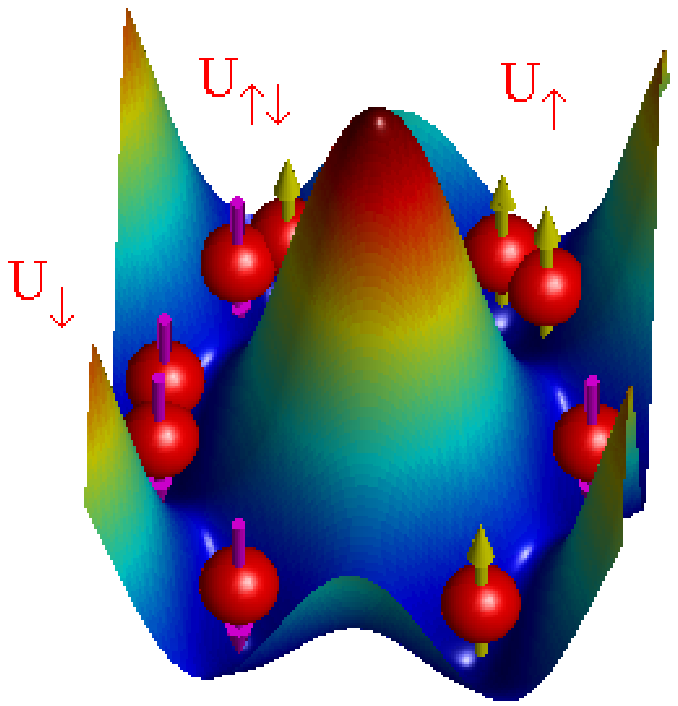}
\end{minipage}
\caption{\!(Color online)  The upper plot shows the schematic diagram of the formation of the bilayer hexagonal optical lattices. The yellow and cyan planes denote the two monolayers of the hexagonal optical lattices which were formed by interacting of three standing wave lasers. The two big arrows show the two vertical standing waves used to couple the two monolayers. The blue line denotes the intra-plane exchange coupling $t_\sigma$ and the blue doublearrow denotes the inter-plane exchange coupling $t^\prime_\sigma$. The lower left plot shows the top view of monolayer optical lattices. In the lower right plot, the yellow uparrow and magenta downarrow denote the two relevant internal states of one atom. $U_{\uparrow}$ and $U_{\downarrow}$ correspond to the strength of the on site interaction energies for the two cold atoms that have the same states $\uparrow$ and $\downarrow$ respectively. $U_{\uparrow\, \downarrow}$ is the interaction strength for the two cold atoms that have the different states $\uparrow$ and $\downarrow$.
\label{fig:1}}
\end{figure}

\textit{Introduction.---}
In recent years, ultracold atoms in optical lattices have been widely used to simulate many-body phenomena in a highly controllable environment \cite{M. Greiner, N. Gemelke, S. Trotzky, C. Orzel, L. Hackermuller, M. Greiner1}. By designing configuration of the atomic system, one can simulate effective theories of the forefront of condensed-matter physics. Recent researches show that graphene has fascinating effects and exhibits particularly rich quantum phases \cite{A. K. Geim, J. P. Reed, K. S. Novoselov, Z. Y. Meng, S. Ladak, A. H. Castro Neto}. Compared to the solid materials, cold atoms in optical lattices are much more controllable. Very recently, the realization of the tunable spin-dependent hexagonal lattices \cite{P. Soltan-Panahi,P. Soltan-Panahi1} indicates that it can be used to investigate the quantum phases of many systems which have a hexagonal geometry \cite{Y. H. Chen,W. Wu,W. Wu1}.\\ \indent One of the goals of studying the cold atoms in optical lattice is to search for the novel states. Supersolid phase (SS) is an exotic state, where the diagonal and off-diagonal long range order coexist. \!Although this novel state still not be found in experiments \cite{E. Kim,E. Kim1,A. Cho,S. Balibar,E. Kim2}, its presence has been confirmed theoretically in lattice model, including bosonic \cite{G. G. Batrouni, F. Hebert,P. Sengupta,S. Diehl, D. Heidarian,S. Wessel,F. Wang,S. Saccani, R. G. Melko} and spin systems \cite{L. Seabra,K. K. Ng, N. Laflorencie, P. Sengupta1,P. Sengupta2}. \!As a powerful tool to tailor the quantum phases, it is reasonable to ask whether the SS phase can be realized in optical lattices. Comparing to the monolayer system, the bilayer one shows fascinating different properties. \!For example, the bilayer graphene has been used to investigate the spin phase transition and the canted antiferromagnetic phase of the $\nu\!=\!0$ quantum hall state \cite{P. Maher,M. Kharitonov,S. Kim,Y. Zhao}. \!Can we design the tunable bilayer hexagonal optical lattices bas-\linebreak ing on the monolayer lattice to search for novel phases?\\ \indent In this Letter, we propose a scheme to investigate the quantum phase transition of cold atoms in bilayer hexagonal optical lattices. As shown in Fig.\,\ref{fig:1}, the bilayer hexagonal optical lattices can be formed by coupling two monolayer hexagonal lattices with two vertical standing waves lasers, where one standing wave has the twice period of the other. The monolayer lattices can be set up by intersection of three laser waves at an angle of $120^\circ$. Using stochastic series expansion (SSE) quantum Monte Carlo (QMC) method \cite{O. F. Syljuasen}, we calculate the ground state phase diagram of cold atoms in this bilayer optical lattices. Our results show that the SS can be realized by adjusting the lattice anisotropy and the ratio of the intra- to inter-layer tunneling. We give a protocol of the observation of the quantum phase transition in real experiments.

\textit{The bilayer hexagonal optical lattices.---}
 Considering cold bosonic atoms (such as $^{87}$Rb) in an optical lattices, we assume that the atoms have two relevant internal states ($|F=1, m_F\!=\!\pm1\rangle$ for $^{87}$Rb) to participate in the dynamics, which are denoted by the two spin index $\sigma=\uparrow,\downarrow$. The atoms are trapped with spin-dependent standing wave laser beams through polarization selection. The three dimensional hexagonal lattices can be set up by intersection of three coplanar laser beams under an angle of $120^\circ$ between each other in the x-y plane and two intersecting waves along the z direction. The total potential of the lattice is
$U(x,y,z) =\sum_{i=1,2,3}V_{\sigma}\sin^2[k_L(x\cos\theta_i+y\sin\theta_i)+\pi/2]+V_{z1\sigma}\cos^2(k_{L_1}z)
+V_{z2\sigma}\cos^2(k_{L_2}z)$, where $\theta_1=\pi/3$, $\theta_2=2\pi/3$, $\theta_3=0$ and $k_L=2\pi/\lambda$ is the optical wave vector with $\lambda=830\,nm$. $V_{\sigma}$ is the barrier height of a single standing wave laser field in the x-y plane \cite{L. M. Duan,S. L. Zhu}, $V_{z1\sigma}$ and $V_{z2\sigma}$ are the barrier heights along z axis, $k_{L_1}=2\pi/\lambda_1$ and $k_{L_2}=2\pi/\lambda_2$ with $\lambda_1=765\,nm$ and $\lambda_2=1530\,nm$ \cite{S. Trotzky}. To form decoupled bilayer lattices, $4V_{z2\sigma} > V_{z1\sigma}$ is required and the ratio of inter- to intra-bilayer potential heights along vertical direction should be so large that the tunneling between two neighbor bilayers can be prevented.
\begin{figure}
\centering
\begin{minipage}[h]{0.5\textwidth}
\centering
\includegraphics[width=2.8in]{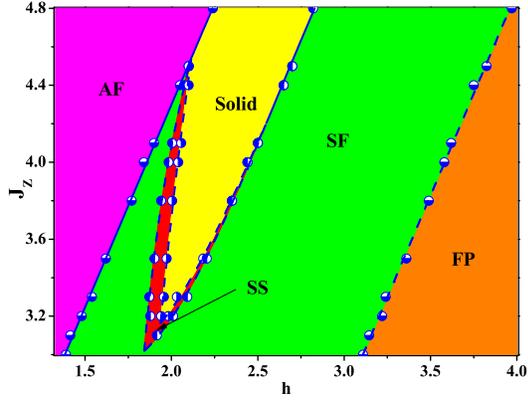}
\end{minipage}
\caption{\!(Color online)  Ground state phase diagram of the intra-layer anisotropy $J_z$ versus effective Zeeman field $h$ for Eq.\,(\ref{eq:1}) with the ratio of intra- to inter-layer tunneling $J\!=\!0.3$. Solid (dashed) lines denote the boundaries of the first (second) order phase transition. The phase diagram contains an antiferromagnet\,(AF), a superfluid\,(SF), a solid, a supersolid\,(SS) and fully polarized state\,(FP).
\label{fig:2}}
\end{figure}

 \indent  Around each minima of the potential the harmonic approximation has been taken and the cold atoms in bilayer optical lattices can be described by a Bose-Hubbard model
\begin{eqnarray}
H &=& -\sum_{\alpha\langle i,j \rangle}(t_{\sigma}a_{\alpha i\sigma}^\dagger a_{\alpha j\sigma}+H.c.)-\sum_{i\sigma}(t^\prime_{\sigma}a_{1i\sigma}^\dagger a_{2i\sigma}+H.c.)\nonumber \\ &&+\frac{1}{2}\sum_{\alpha i\sigma}[U_\sigma n_{\alpha i\sigma}(n_{\alpha i\sigma}-1)]+U_{\uparrow\downarrow}\sum_{\alpha i}n_{\alpha i\uparrow}n_{\alpha i\downarrow}\nonumber\\ &&-\mu\sum_{\alpha i}(n_{\alpha i\uparrow}-n_{\alpha i\downarrow})\, ,
\label{eq:1}
\end{eqnarray}
where $\alpha $ denotes the layer index 1, 2, $a_{i\sigma}^\dagger\, (a_{i\sigma})$ is the creation (annihilation) operator of the bosonic atom at site i for spin $\sigma$, $n_{i\sigma}=a_{i\sigma}^\dagger a_{i\sigma}$ is the occupation number on site i. $\langle i,j \rangle$ runs over nearest neighbors. We will only focus on the regime of strong coupling, $U_\sigma,\,U_{\uparrow\downarrow}\gg t_{\sigma},\,t^\prime_{\sigma}$, i.e., the mott insulator regime, where each lattice site is occupied by one atom.
\begin{figure}
\centering
\begin{minipage}[t]{0.5\textwidth}
\centering
\includegraphics[width=2.8in]{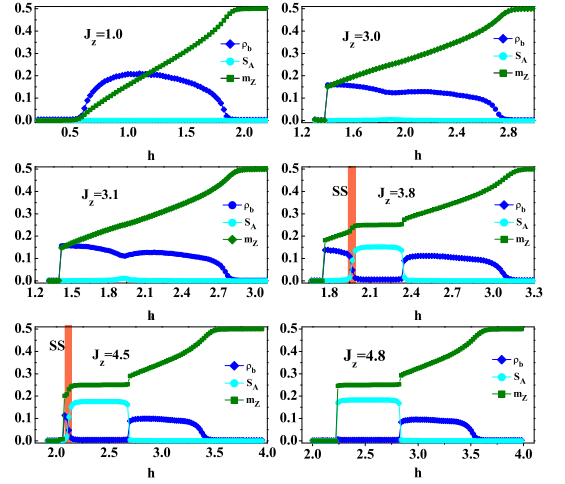}
\end{minipage}
\caption{\!(Color online)  The superfluid density $ \rho_b $, solid order parameter $ S_A $ and magnetization $ m_z $ versus effective Zeeman field $ h $ for different intra-layer anisotropy $ J_z $, where the ratio of intra- to inter-layer tunneling $J=0.3$.
\label{fig:3}}
\end{figure}

\textit{The phase diagram of ground state.---} We define the reduced interaction parameters as $J_\perp=2t^\prime_\uparrow t^\prime_\downarrow /U_{\uparrow\downarrow}$, $J=2t_\uparrow t_\downarrow /U_{\uparrow\downarrow}$, $J_z=-({t_\uparrow}^2 + {t_\downarrow}^2)/(t_\uparrow t_\downarrow)+2U_{\uparrow\downarrow}[{t_\uparrow}/(U_{\uparrow} t_\downarrow)+{t_\downarrow}/(U_{\downarrow}t_\uparrow)]$ and the effective Zeeman field $h=2\mu-2({t^\prime_\uparrow}^2+{t_\uparrow}^2)/U_{\uparrow}+2({t^\prime_\downarrow}^2+{t_\downarrow}^2)/U_{\downarrow}$. In the following, we will take $J_\perp=1$ as energy unit and use the stochastic series expansion quantum Monte Carlo method to calculate the ground state phase diagram of Eq.\,(\ref{eq:1}). In our simulation, periodic boundary condition is imposed and the temperature is set inversely proportional to $2L$ with $L$\,=\,8, 10, 12 and 16. The lowest temperature can reach to 0.03125, which is low enough to insure that we can investigate the ground state properties.\\ \indent We focus on the low energy physics of the system where $J\ll 1$ is required (in this Letter, $ 0.1  \le J \le 0.4 $). The four eigenstates of two atoms coupled by the inter-layer tunneling i.e., singlet state $ ( | s \rangle=\frac{1}{\sqrt2}(|\!\uparrow\downarrow\rangle-|\!\downarrow\uparrow\rangle )) $  and triplet states  $ ( |t_0\rangle=\frac{1}{\sqrt2}(|\!\uparrow\downarrow\rangle+|\!\downarrow\uparrow\rangle ), |t_1\rangle=|\!\uparrow\uparrow\rangle, |t_{-1}\rangle=|\!\downarrow\downarrow\rangle) $, are conveniently used to describe the ground state of the system. When there exist a weak effective Zeeman field, the triplets split into three and the lowest state $ ( |t_1\rangle )$ keeps dropping with the effective Zeeman field increasing. It is clear that the contribution of $|t_0\rangle$ and $|t_{-1}\rangle$ to the ground state is negligible. When the effective Zeeman field exceeds a critical value, $ ( |t_1\rangle )$ take the place of $|s\rangle$ to be the lowest energy level and the system will be a fully polarized state. To describe the physics of the system, a bosonic operators $ b^\dagger_i=(-1)^i/\sqrt {2} (s^+_{1,i}-s^+_{2,i}) $ can be defined and it transforms $|s\rangle$ into $ ( |t_1\rangle )$ \cite {K. K. Ng}. It has the following operation: $b^\dagger_i \,|t_{-1}\rangle_i=| s \rangle_i$, $b^\dagger_i\,|s \rangle_i=| t_{1} \rangle_i$ and $b^\dagger_i\, |t_{1}\rangle_i=0$.\\ \indent To character the superfluid state, we compute the condensate density $ \rho_b=1/N^2\sum_{i,j}\langle b_i^\dagger b_j\rangle$ ($N$ is the number of inter-layer bond, $N=2\times L^2$). For solid state, the usually static structure parameter $ S(\mathbf {Q})$ is not suitable for the honeycomb lattice and we define the structure parameter
$S_A=1/N^2\sum _{\langle i,j\rangle}(\epsilon _i\epsilon_j\langle n_in_j\rangle)$,
where $ \epsilon_i=+1\:(-1) $ for i on sublattice $ A \:(B)$, $ n_i $ represents the number of $ ( |t_1\rangle )$ state on dimmer i. The antiferromagnet is characterized by the staggered magnetization $ M^s_z=1/(2N)\langle \sum_i\epsilon_i(S^z_{1i}-S^z_{2i})\rangle $. The magnetization is calculated by $ M_z=1/(2N)\langle \sum_i(S^z_{1i}+S^z_{2i})\rangle $.
\begin{figure}
\centering
\begin{minipage}[t]{0.5\textwidth}
\centering
\includegraphics[width=2.4in]{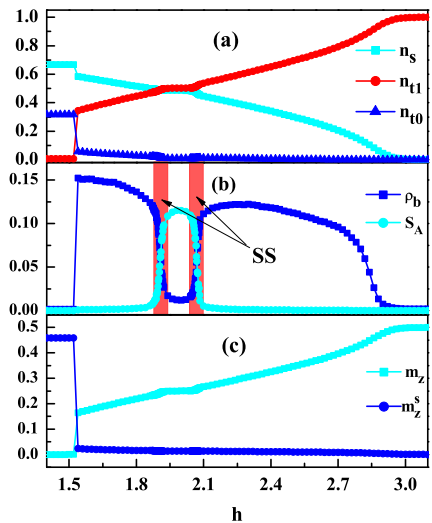}
\end{minipage}
\caption{\!(Color online) (a) The number densities of the states: $ |s\rangle=\frac{1}{\sqrt2}(|\!\uparrow\downarrow\rangle-|\!\downarrow\uparrow\rangle)$, $ |t_1\rangle=|\!\uparrow\uparrow\rangle $ and $ |t_0\rangle=\frac{1}{\sqrt2}(|\!\uparrow\downarrow\rangle+|\!\downarrow\uparrow\rangle$ versus effective Zeeman field $ h $. (b) The superfluid density $ \rho_b $ and solid order parameter $ S_A $ versus effective Zeeman field $ h $. (c) The magnetization $ m_z $ and the staggered magnetization $ m_z^s $ versus effective Zeeman field $ h $. Here, the intra-layer anisotropy $J_z =3.3$ and the ratio of intra- to inter-layer tunneling $J=0.3$.
\label{fig:4}}
\end{figure}

\begin{figure}
\centering
\begin{minipage}[b]{0.5\textwidth}
\centering
\includegraphics[width=3.0in]{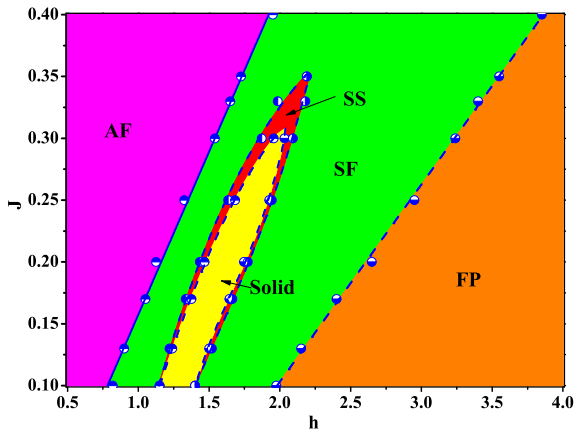}
\end{minipage}
\caption{\!(Color online)  Ground state phase diagram of the ratio of intra- to inter-layer tunneling $J$ versus effective Zeeman field $h$ for Eq.\,(\ref{eq:1}) with the intra-layer anisotropy $J_z\!=\!3.3$. Solid (dashed) lines denote the boundaries of the first (second) order phase transition. The phase diagram contains an antiferromagnet\,(AF), a superfluid\,(SF), a solid, a supersolid\,(SS) and fully polarized state\,(FP).
\label{fig:6}}
\end{figure}

 The ground state phase diagram of $ H $ in the $J_z-h$ plane with $J=0.3$ is shown in Fig.\,\ref {fig:2}, where $J_{z}$ is the intra-layer anisotropy of interactions and $h$ is the effective Zeeman field case. Order parameters $\rho_b$, $S_A$ and $m_z$ vs the effective Zeeman field $h$ for different $ J_z $ are shown in Fig.\,\ref {fig:3}. When $J_z=1$, i.e., the isotropic case, the system is in the SF state, which is characterized by $\rho_b>0$ and $S_A=0$. In this state, the U(1) symmetry of the system is broken by the Bose-Einstein condensates of $|t_1\rangle$. When $J_z>1$, without loss of generality, we select $J_z =3.3$, the evolution of the number densities and order parameters with the effective Zeeman field $h$ increasing is shown in Fig.\,\ref {fig:4}. At lower effective Zeeman field, the only nonvanishing order parameter $ M^s_z $ shows that the system is in the antiferromagnet state. This state is the traditional Ising order and the system has the staggered up and down arrangement in both intra- and inter-layer. When the effective Zeeman field increased to a critical value, the staggered magnetization drops abruptly while the magnetization has the inverse behavior, this denotes the first order transition from AF to SF. Between $1.0 <J_z< 3.0$, with effective Zeeman field increasing, the system only experiences AF, SF and FP states.
\begin{figure}
\centering
\begin{minipage}[t]{0.5\textwidth}
\centering
\includegraphics[width=2.2in]{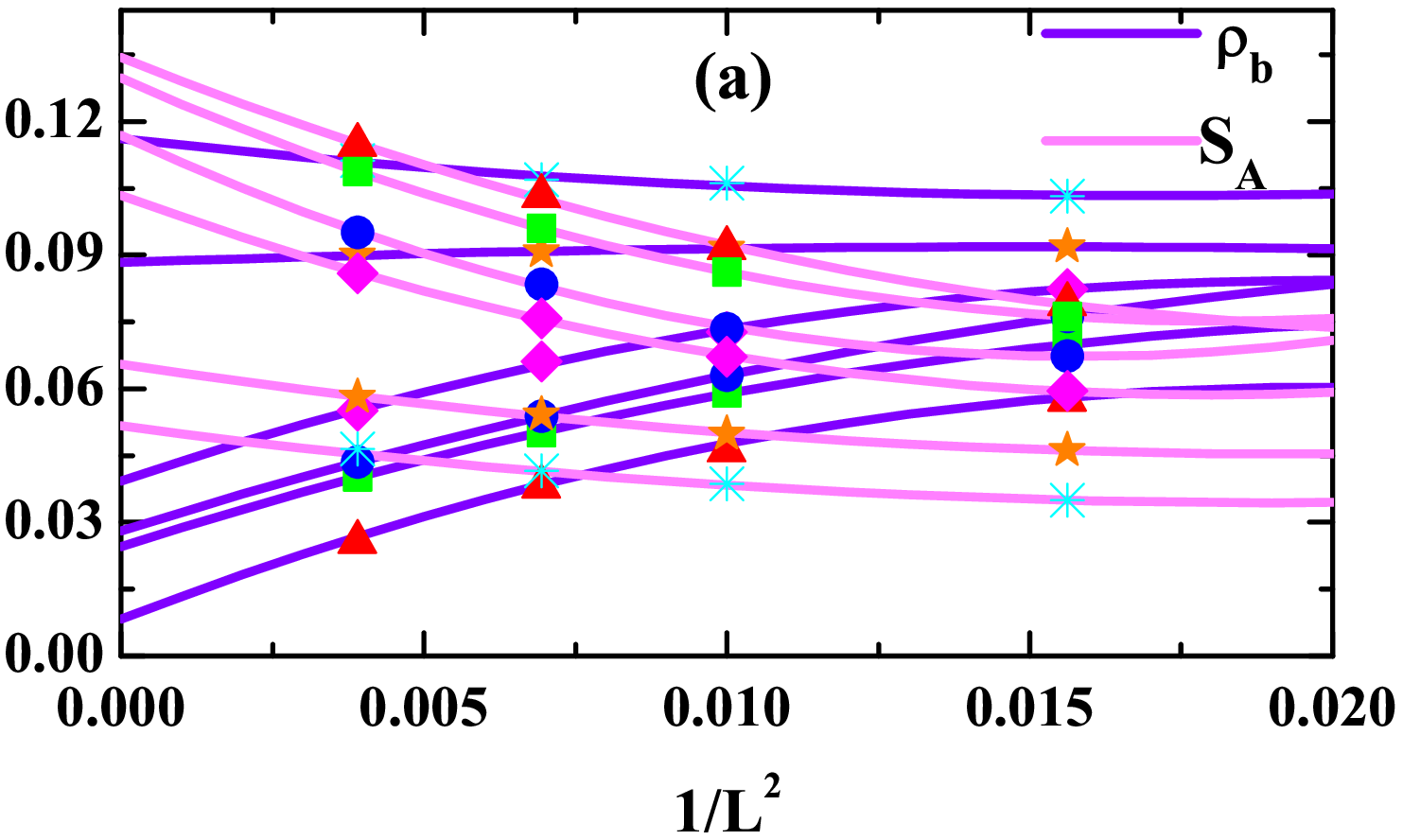}\\
\includegraphics[width=2.2in]{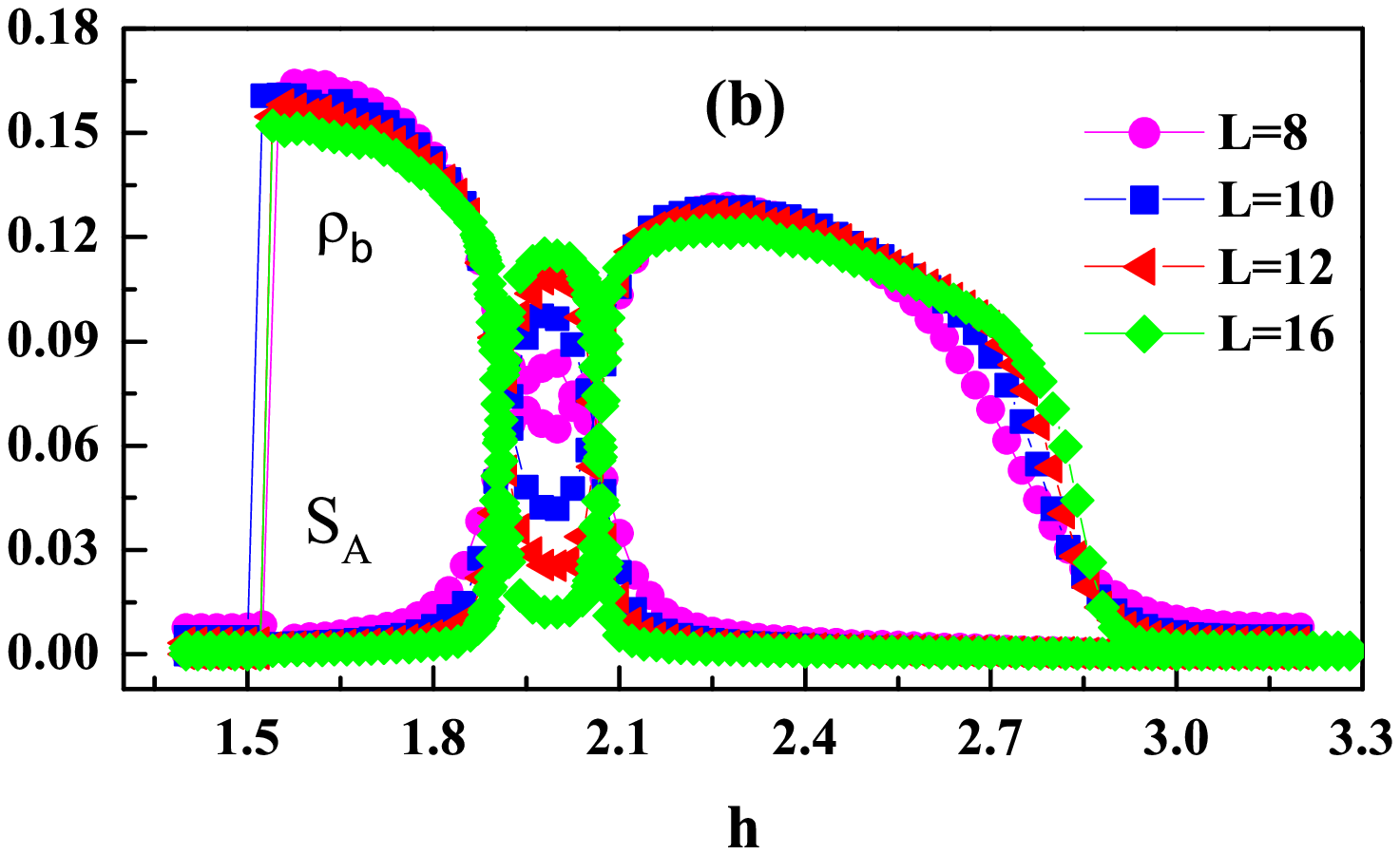}
\end{minipage}
\caption{\!(Color online)  (a) Finite size extrapolation of the superfluid density $\rho_b$ and the solid order parameter $S_A$. Effective Zeeman field $h$ changes from 2.0 to 2.1 with increments 0.02 from bottom to top for $\rho_b$ and from top to bottom for $S_A$. (b) The results of $\rho_b$ and $S_A$ for $L$=8, 10, 12 and 16, where the intra-layer anisotropy $J_z =3.3$ and the ratio of intra- to inter-layer tunneling $J=0.3$.
\label{fig:5}}
\end{figure}
\\ \indent As the anisotropy $J_{z}>3.0$, see Fig.\,\ref {fig:4}, with the effective Zeeman field increasing, the number density of $|t_1\rangle$ jumps abruptly to a finite value in the transition from AF to SF. Then, the triplet state $|t_1\rangle$ keeps increasing to half filled and the system is in a solid state which is made of staggered singlet state and triplet state $|t_1\rangle$. This state is characterized by $S_A > 0$ and $\rho_b=0$ and its obvious feature is the plateau in the $m_z$ curve. After the platform, $n_{t1}$ keeps on increasing and the SF state reappear. In the intermediate region between SF and solid or between solid and SF, there appears a supersolid state which is characterized by both $S_A > 0$ and $\rho_b>0$ in the thermodynamic limit. Finite size extrapolation results shown in Fig.\,\ref {fig:5} indicate that the SS is nonzero when $L$ is extrapolated to infinite. It is obviously that the transitions of SF-SS, solid-SS, SS-solid and SF-FP are second order. When $J_z=4.8$, the SS state disappears, the transitions of AF-solid and solid-SF are first order. \\ \indent In the following, we want to consider the ground state phase diagram in the ratio of intra- to inter-layer tunneling versus effective Zeeman field case, i.e., $J-h$ plane at fixed $J_z$. The phase diagram with $J_z=3.3$ is shown in Fig.\,\ref {fig:6}. In the small $J$, the system only appears AF, SF, solid and FP. When $J$ increases to nearly 0.15, two sharp regions of SS appear between SF and QS and they turn wider with $J$ increasing and merger into one at $J=0.3$, where QS phase disappears. Within $0.17<J<0.32$ the SS state is stable in the thermodynamic limit and it vanishes and leaves SF phase only when $J$ up to 0.35.

\textit{Experimental protocol.---}
The experimental protocol of cold atoms in the bilayer hexagonal optical lattices can be taken as follows: $^{87}$Rb Bose-Einstein condensates can be created up to $10^6$ atoms in the two states ($|\!\uparrow\rangle\equiv |F=1, m_F\!=\!1\rangle$, $|\!\downarrow\rangle\equiv |F=1, m_F\!=\!-1\rangle$). Here $F$ denotes the total angular momentum and $m_F$ the magnetic quantum number of the state. To form the hexagonal lattices, three optical standing waves were aligned intersection under an angle of $120^\circ$ in the x-y plane. The laser beams can be produced by a Ti:sapphire laser operated at a wavelength $\lambda=830\,nm$ (red detuned) \cite{P. Soltan-Panahi,P. Soltan-Panahi1}. To form the bilayer structure, orthogonal to the x-y plane, two intersecting standing waves are settled. The light for the two standing waves can be created by a 1,530 nm fibre laser and a Ti:sapphire laser running at 765 nm \cite{S. Folling}. For $^{87}$Rb atom, the energy scale of the typical tunneling rate $t_\sigma/\hbar$ and $t_\sigma^\prime/\hbar$ can be chosen from 0 to a few kH$_Z$, the on-site interaction $U_\sigma/\hbar$ and $U_{\uparrow\downarrow}/\hbar$ can be a few kH$_Z$ at zero magnetic field or much more larger near the Feshbach resonance. We can easily choose $t_\sigma/\hbar$, $t_\sigma^\prime/\hbar$, $U_\sigma/\hbar$ and $U_{\uparrow\downarrow}/\hbar$ to satisfy $\frac{1}{10}$ $<$ $\frac{t_\sigma(t_\sigma^\prime)}{U_\sigma(U_{\uparrow\downarrow})}$ $<$ $\frac{1}{30}$, where the system is in the Mott insulating area. \\ \indent To detect the magnetization $m_z$ and staggered magnetization $m_z^s$, a quantum polarization spectroscopy method can be used. In this method, a pulsed polarized light was sent to the atoms trapped in the lattice, after the light polarization coupled with the atomic spin, the atoms magnetization can be detected by the light polarization rotations \cite{K. Eckert, G. De Chiara}. The superfluid density $\rho_b$ in a range 0 $\sim$ 0.25 could be obtained from spin-spin correlation which can be detected by the noise correlation method \cite{S. Folling1,E. Altman,V. W. Scarola}. Spin structure factor $S_A$ in a range 0 $\sim$ 0.2 can be detected by the Bragg scattering. The spatial correlation function can be directly measured by using spatially correlated imaging light \cite{T. A. Corcovilos, J. S. Douglas}.

\textit{Conclusion.---}
In summary, we propose a scheme to investigate the quantum phase transition of cold atoms in the bilayer hexagonal optical lattices. This bilayer optical lattices can be realized by coupling two monolayer optical lattices with two intersecting standing waves. Our results show that the phase diagrams of this system contain an antiferromagnet, a solid, a superfluid, a fully polarized state and, especially, a supersolid. Thus, the cold atoms in bilayer hexagonal optical lattices are a practicable way to tailor quantum phases and could be used to search for the novel states in real experiments.

This work was supported by the NKBRSFC under grants Nos. 2011CB921502, 2012CB821305, 2009CB930701, 2010CB922904, NSFC under grants Nos. 10934010, 11228409, 61227902 and NSFC-RGC under grants Nos. 11061160490 and 1386-N-HKU748/10.


\begin{thebibliography}{99}

\bibitem{M. Greiner} M. Greiner, O. Mandel, T. Esslinger, T. W. H\"ansch and I. Bloch, Nature \textbf{415}, 39 (2002).

\bibitem{N. Gemelke} N. Gemelke, X. B. Zhang, C. L. Hung and C. Chin, Nature \textbf{460}, 995 (2009).


\bibitem{S. Trotzky} S. Trotzky, P. Cheinet, S. F\"olling, M. Feld, U. Schnorrberger, A. M. Rey, A. Polkovnikov, E. A. Demler, M. D. Lukin and I. Bloch, Science \textbf{319}, 295 (2008).

\bibitem{C. Orzel} C. Orzel, A. K. Tuchman, M. L. Fenselau, M. Yasuda, and M. A. Kasevich, Science \textbf{291}, 2386 (2001).

\bibitem{L. Hackermuller} L. Hackerm\"uller, U. Schneider, M. Moreno-Cardoner, T. Kitagawa, T. Best,
S. Will, E. Demler, E. Altman, I. Bloch and B. Paredes, Science \textbf{327}, 1621 (2010).


\bibitem{M. Greiner1} M. Greiner, I. Bloch, O. Mandel, T. W. H\"ansch and T. Esslinger, Phys. Rev. Lett. \textbf{87}, 160405 (2001).

\bibitem{A. K. Geim} A. K. Geim and K. S. Novoselov, Nature Mate. \textbf{6}, 183 (2007).

\bibitem{J. P. Reed} J. P. Reed, B. Uchoa,  Y. I. Joe, Y. Gan, D. Casa, E. Fradkin and P. Abbamonte, Science \textbf{330}, 805 (2010).

\bibitem{K. S. Novoselov} K. S. Novoselov, A. K. Geim, S. V. Morozov, D. Jiang, M. I. Katsnelson, I. V. Grigorieva, S. V. Dubonos and A. A. Firsov, Nature \textbf{438}, 197 (2005).

\bibitem{Z. Y. Meng} Z. Y. Meng, T. C. Lang, S. Wessel, F. F. Assaad and A. Muramatsu, Nature \textbf{464}, 847 (2010).

\bibitem{S. Ladak} S. Ladak, D. E. Read, G. K. Perkins, L. F. Cohen and W. R. Branford, Nat. Phys. \textbf{6}, 359 (2010).

\bibitem{A. H. Castro Neto} A. H. Castro Neto, F. Guinea, N. M. R. Peres, K. S. Novoselov and A. K. Geim, Rev. Mod. Phys. \textbf{81}, 109 (2009).


\bibitem{P. Soltan-Panahi} P. Soltan-Panahi, J. Struck, P. Hauke, A. Bick, W. Plenkers, G. Meineke, C. Becker, P. Windpassinger, M. Lewenstein and K. Sengstock, Nat. Phys. \textbf{7}, 434 (2011).
 \bibitem{P. Soltan-Panahi1} P. Soltan-Panahi, D. S. L\"uhmann, J. Struck, P. Windpassinger and K. Sengstock, Nat. Phys. \textbf{8}, 71 (2012).

\bibitem{Y. H. Chen} Y. H. Chen, H. S. Tao, D. X. Yao and W. M. Liu, Phys. Rev. Lett. \textbf{108}, 246402 (2012).
\bibitem{W. Wu}  W. Wu, S. Rachel, W. M. Liu and K. L. Hur, Phys. Rev. B \textbf{85}, 205102 (2012).
\bibitem{W. Wu1}  W. Wu, Y. H. Chen, H. S. Tao, N. H. Tong, W. M. Liu, Phys. Rev. B \textbf{82}, 245102 (2010).

\bibitem{E. Kim} E. Kim and M. H. W. Chan, Nature \textbf{427}, 225 (2004).
\bibitem{E. Kim1} E. Kim and M. H. W. Chan, Science \textbf{305}, 1941 (2004).
\bibitem{A. Cho} A. Cho, Science \textbf{336}, 661 (2012).
\bibitem{S. Balibar} S. Balibar, Nature \textbf{464}, 176 (2010).
\bibitem{E. Kim2} E. Kim and M. H. W. Chan, Phys. Rev. Lett. \textbf{109}, 155301 (2012).

\bibitem{G. G. Batrouni} G. G. Batrouni and R. T. Scalettar,
Phys. Rev. Lett. \textbf{84}, 1599 (2000).
\bibitem{F. Hebert} F. H\'ebert, G. G. Batrouni,  R. T. Scalettar, G. Schmid, M. Troyer and A. Dorneich,
Phys. Rev. Lett. \textbf{65}, 014513 (2001).
\bibitem{P. Sengupta} P. Sengupta, L. P. Pryadko, F. Alet, M. Troyer and G. Schmid,
Phys. Rev. Lett. \textbf{94}, 207202 (2005).
\bibitem{S. Diehl} S. Diehl, M. Baranov, A. J. Daley and P. Zoller, Phys. Rev. Lett. \textbf{104}, 165301 (2010).
\bibitem{D. Heidarian} D. Heidarian and A. Paramekanti, Phys. Rev. Lett. \textbf{104}, 015301 (2010).
\bibitem{S. Wessel} S. Wessel and M. Troyer, Phys. Rev. Lett. \textbf{95}, 127205 (2005).
\bibitem{F. Wang} F. Wang, F. Pollmann and A. Vishwanath, Phys. Rev. Lett. \textbf{102}, 017203 (2009).
\bibitem{R. G. Melko} R. G. Melko, A. Paramekanti, A. A. Burkov, A. Vishwanath, D. N. Sheng and L. Balents, Phys. Rev. Lett. \textbf{95}, 127207 (2005).
\bibitem{S. Saccani} S. Saccani, S. Moroni and M. Boninsegni, Phys. Rev. Lett. \textbf{108}, 175301 (2012).


\bibitem{L. Seabra} L. Seabra and N. Shannon, Phys. Rev. Lett. \textbf{104}, 237205 (2010).
\bibitem{K. K. Ng} K. K. Ng and T. K. Lee, Phys. Rev. Lett. \textbf{97}, 127204 (2006).
\bibitem{N. Laflorencie} N. Laflorencie and F. Mila, Phys. Rev. Lett. \textbf{99}, 027202 (2007).
\bibitem{P. Sengupta1} P. Sengupta and C. D. Batista, Phys. Rev. Lett. \textbf{98}, 227201 (2007).
\bibitem{P. Sengupta2} P. Sengupta and C. D. Batista, Phys. Rev. Lett. \textbf{99}, 217205 (2007).

\bibitem{P. Maher} P. Maher, C. R. Dean, A. F. Young, T. Taniguchi, K. Watanabe, K. L. Shepard, J. Hone and P. Kim, Nat. Phys. \textbf{9}, 154 (2013).
\bibitem{M. Kharitonov} M. Kharitonov, Phys. Rev. Lett. \textbf{109}, 046803 (2012).
\bibitem{S. Kim} S. Kim, K. Lee and E. Tutuc, Phys. Rev. Lett. \textbf{107}, 016803 (2011).
\bibitem{Y. Zhao} Y. Zhao, P. Cadden-Zimansky, Z. Jiang and P. Kim, Phys. Rev. Lett. \textbf{104}, 066801 (2010).


\bibitem{O. F. Syljuasen} O. F. Sylju\r{a}sen and A. W. Sandvik, Phys. Rev. E \textbf{66}, 046701 (2002).

\bibitem{L. M. Duan} L. M. Duan, E. Demler and M. D. Lukin,
Phys. Rev. Lett. \textbf{91}, 090402 (2007).
\bibitem{S. L. Zhu} S. L. Zhu, B. Wang and L. M. Duan,
Phys. Rev. Lett. \textbf{98}, 260402 (2007).


\bibitem{S. Folling} S. F\"olling, S. Trotzky, P. Cheinet, M. Feld, R. Saers, A. Widera, T.Mu\..ller and I. Bloch, Nature \textbf{448}, 1029 (2007).

\bibitem{K. Eckert} K. Eckert, O. Romero-isart, M. Rodriguez, M. Lewenstein, E. S. Polzik and A. Sanpera, Nat. Phys. \textbf{4}, 50 (2008).
\bibitem{G. De Chiara} G. De Chiara, O. Romero-Isart and A. Sanpera, Phys. Rev. A \textbf{83}, 021604(R) (2011).

\bibitem{S. Folling1} S. F\"olling, F. Gerbier, A. Widera, O. Mandel, T. Gericke and I. Bloch, Nature \textbf{434}, 481 (2005).
\bibitem{E. Altman} E. Altman, E. Demler and M. D. Lukin, Phys. Rev. A \textbf{70}, 013603 (2004).
\bibitem{V. W. Scarola} V. W. Scarola, E. Demler and S. DasSarma, Phys. Rev. A \textbf{73}, 051601(R) (2006).

\bibitem{T. A. Corcovilos} T. A. Corcovilos, S. K. Baur, J. M. Hitchcock, E. J. Mueller and R. G. Hulet, Phys. Rev. A \textbf{81}, 013415 (2010).

\bibitem{J. S. Douglas} J. S. Douglas and K. Burnett, Phys. Rev. A \textbf{82}, 033434 (2010).



\end{thebibliography}
\end{document}